\documentstyle[12pt,aasms4]{article} 
\def\etal{{et~al.}\ }
\def\vol#1  {{{#1}{\rm,}\ }}

\def\lya{{\rm Ly}\alpha}

\def\etal{et al.\ }

\def\clock{\count0=\time \divide\count0 by 60
     \count1=\count0 \multiply\count1 by -60 \advance\count1 by \time
     \number\count0:\ifnum\count1<10{0\number\count1}\else\number\count1\fi}

\begin{document}
\title{On the Cluster Sunyaev-Zel'dovich Effect and Hubble Constant}
\author{Renyue Cen}
\vskip 0.5cm
\centerline{Princeton University Observatory, Princeton, NJ 08544}
\vskip 0.5cm
\centerline{cen@astro.princeton.edu}
\vskip 0.7cm
%\centerline{draft of Sept. 28, 1997}

\begin{abstract}

This study shows one important effect 
of preexistent cosmic microwave
background temperature fluctuations
on the determination
of the Hubble constant through Sunyaev-Zel'dovich effect of
clusters of galaxies,
especially when coupled with the gravitational lensing effect by 
the same clusters.
The effect results in
a broad distribution of the apparent Hubble constant.
The combination of this effect with other systematic effects
such as the Loeb-Refregier Effect seems to provide an
explanation for the observationally derived values of the Hubble 
constant
currently available based on
the Sunyaev-Zel'dovich effect, if the true value of the Hubble constant
is $60-80~$km/s/Mpc.
It thus becomes possible that the values of the Hubble constant measured
by other techniques which generally give a value around $60-80~$km/s/Mpc
be reconciled with the SZ effect determined values of the Hubble constant,
where are systematically lower than others
and have a broad distribution.

\end{abstract}


\section{Introduction}

The Sunyaev-Zel'dovich (SZ) effect 
of a cluster of galaxies on the cosmic microwave background (CMB)
photons can be used to determine the distance
to the cluster hence the Hubble constant ($H_0$),
when analysed in conjunction with X-ray observations of the cluster
(Cavaliere, Danese, \& De Zotti 1977; Gunn 1978;
Silk \& White 1978;
Birkinshaw 1979). 
For an excellent recent review on 
this subject and other SZ related topics, see  Rephaeli (1995 and
references therein).
The accuracy of the Hubble constant
determination depends upon the accuracy of
several assumptions involving both sets of observations (radio and X-ray).
Perhaps among the most important are the assumptions
of sphericity, isothermality of clusters of galaxies
(e.g., Inagaki \etal 1995). In this {\it Letter} 
we point out a completely separate effect 
on the determination of the Hubble constant
due to preexistent, small-amplitude
CMB temperature fluctuations before the photons undergo 
the SZ effect through a cluster.
The effect is significantly amplified by the gravitational lensing
of the CMB photons by the cluster,
because the SZ observational beam size 
is typically comparable to Einstein radius of the source-lens system.
This effect, when coupled with some systematic effects such as
the one proposed by Loeb \& Refregier (1997) due to the systematic
over-removal of background point radio sources in the beam,
may provide an explanation for the observed distribution of $H_0$ 
determined by SZ effect.

\section{CMB Fluctuations and $H_0$ Determination Using SZ Effect}

The detections of CMB temperature fluctuations
on arcminute scales are mostly upper limits or marginal
(e.g., Partridge \etal 1997),
primarily because of limited 
sky coverages and instrumental sensitivities.
However, there are a number of 
physical mechanisms suggested, which
should generate appreciable fluctuations on the relevant scales.
For example,
Persi \etal (1995)
(see also Scaramella, Cen, \& Ostriker 1993)
show that SZ effect due to non-cluster gas
can generate $\Delta T/T\sim 10^{-6}-10^{-5}$ on the arcminute scale,
produced naturally by shock heated gas in a network
of filaments and sheets during the phase of gravitational collapse
of large-scale structure.
Loeb (1996) shows that
one would expect $\Delta T/T\sim 10^{-6}-10^{-5}$
due to bremsstrahlung emission from $\lya$ clouds,
given the observed/required meta-galactic ultraviolet radiation field.
Depending on the universal ionization history,
the Ostriker-Vishniac Effect (Ostriker \& Vishniac 1986; Vishniac 1987)
could also make 
appreciable contributions to the CMB fluctuations on
the relevant scales 
(e.g., Persi \etal 1995).
To be quantitatively definitive, 
we will adopt CMB temperature fluctuations observed
by Partridge \etal (1997) and assume that
they are background CMB temperature fluctuations, meaning that
they exist before 
CMB fluctuations are further induced by hot gas in clusters of galaxies.
This is not to say that 
these fluctuations are primordial; 
all that we need to assume is that 
there exists appreciable CMB fluctuations at redshift higher
than that of the clusters for which we study
their individual SZ effect and then determine the Hubble
constant (when coupled with X-ray observations of the clusters).

To clearly illustrate the point, we will first adopt a set of numbers 
and make a few simplifying assumptions:

\noindent
1) the true Hubble constant is $H_{0,true}=65$km/s/Mpc;

\noindent
2) X-ray observations of clusters and the interpretations of them
   are error-free;

\noindent
3) the clusters are spherical with a singular isothermal profile for
   the total mass with
   one-dimensional velocity dispersion of $1021~$km/s
   (corresponding to a gravitational lensing bending angle of $30^"$),
   and a true SZ effect of $(\Delta T/T)_{SZ}=2.0\times 10^{-4}$
   (the observed arcs by gravitational lensing of rich clusters
   have sizes comparable to $30^"$ and the observed SZ effect in clusters
   is about $(1-6)\times 10^{-4}$ with a mean about $2.0\times 10^{-4}$
   [see Rephaeli 1995 for a summary]);

\noindent
4) the profile of the observational beam is, for the
simplicity of calculation, assumed to 
be a two-dimensional top-hat with a radius of $60^"$
   and the beam is precisely centered on each cluster center;

\noindent
5) we adopt the CMB fluctuations around arcminute scales
   from Partridge \etal (1997), adapted as columns 1 and 2 in Table 1,
   and assume that the fluctuations are Gaussian.
   Note that, when the CMB maps are generated, we follow that 
   the scales indicated in Table 1 are FWHM's but with
   a Gaussian profile (i.e., FWHM$=2.35\sigma_g$,
   where $\sigma_g$ is the radius of the Gaussian window).
   This is not to be confused with the SZ observation's beam
   shape, which is assumed to be a top-hat for the convenience of 
   calculation (assumption 4, above).

The rms fluctuation of the CMB map 
{\it just due to pre-SZ CMB fluctuations alone},
as indicated by the third column in Table 1, 
on a top-hat circle of radius $60^"$ is $6.8\times 10^{-6}$.
The distribution of fluctuations 
on a top-hat circle of radius $60^"$ is
shown as heavy histogram in Panel (a) of Figure 1 for a sample of 1000
random beams.
The light dashed curve is the Gaussian fit with the same variance 
and zero mean.
Since the Hubble constant determined
from SZ effect is 
proportional to $(\Delta T/T)_{SZ}^{-2}$
(Cavaliere, Danese, \& De Zotti 1977; Gunn 1978; Silk \& White 1978;
Birkinshaw 1979), 
we immediately obtain a distribution of $H_0$,
given the distribution of $\Delta T/T$.
Thus, the light dashed curve in Panel (a) of Figure 1, $f(\Delta T/T)$,
can be translated into a distribution of $H_0$, which
is shown as the dashed curve in Figure 2, $g(H_0)$.
The short vertical bar at $H_0=65$km/s/Mpc
indicates the (adopted) true value of the Hubble constant.
Note that, although the light dashed curve in Panel (a)
is Gaussian, the dashed curve in Figure 2 is {\it not} 
Gaussian because the translation from $f(\Delta T/T)$ to 
$g(H_0)$ is {\it nonlinear}.

Next, we examine how gravitational lensing of CMB photons
by clusters alter the CMB fluctuations on the arcminute scale 
around the clusters.
In order to do this, we first generate synthetic CMB maps which 
agree with observations on $6^"-80^"$ scales tabulated in Table 1.
We find that a two-dimensional power-law
power spectrum of index $-0.42$ (i.e., $P_k\propto k^{-0.42}$)
produces a satisfactory fit to observations, as indicated 
by the third column in Table 1.
We normalize the fluctuations by requiring
$(\Delta T/T)_{syn}=(\Delta T/T)_{obs}=1.2\times 10^{-5}$
at FWHM$=60^"$.

For each synthetic CMB map 
of size $12^{'}.8\times 12^{'}.8$, 
with pixel size of $6^"\times 6^"$,
we center the cluster
at the optical axis connecting the observer and the center of the 
source plane map.
For simplicity we assume that we live in an $\Omega=1$ universe,
the sources (CMB photons) are placed at a distance of
$D_s=6000h^{-1}$Mpc (i.e., 
$z_s>>1$) from the observer,
and the lens (the cluster) is at a distance of $D_l=1000h^{-1}$Mpc (i.e, 
$z_l\sim 0.44$) from the observer
(note that only the combination $(D_s-D_l)/D_s$ enters
our calculation of ray tracing; see below).
Each CMB source pixel of 
size $6^"\times 6^"$ is divided into smaller (square) sub-pixels.
The number of sub-pixels ($N_{sub}$)
for each such division
is adaptive
depending on the amplification ($\mu$) 
of the coarse pixels (for an optimal efficiency of calculation).
The photons of sub-pixels are ray traced 
through the cluster potential and collected in the image plane
with pixels of exactly the same size ($6^"\times 6^"$)
as that of the coarse pixels in the source plane.
The accuracy of the ray tracing method is measured as follows.
We keep increasing the number of sub-pixels
until the fluctuations on a top-hat of radius
$60^"$ of the CMB map in the image plane is less than
$5\times 10^{-7}$ for a source plane map with zero fluctuations on 
all scales; note that
an ideal method should give zero fluctuations in the image
plane map in this case.
This is sufficiently accurate since the real fluctuations
are on the order of $1.0\times 10^{-5}$.
We find that $N_{sub}(\mu)=4096\mu^6$ is required to
give such a satisfactory accuracy.

We ray trace 1000 independent maps and the
final distribution of $\Delta T/T$ on the top-hat beam of radius $60^"$,
$F(\Delta T/T)$, is shown as heavy solid histogram in Panel (b) of Figure 1.
The variance of $F(\Delta T/T)$ is $1.3\times 10^{-5}$
compared to $6.8\times 10^{-6}$, the variance of $f(\Delta T/T)$,
the distribution prior to the gravitational lensing effect by the clusters.
The light dashed curve in the same panel is the Gaussian fit
with the same variance ($1.3\times 10^{-5}$).

We can translate the light dashed curve in Panel (b) of Figure 1 
into a distribution of $H_0$, shown as  
the dotted curve in Figure 2.
It is worthwhile to consider
other effects on the determination of SZ effect of clusters.
Loeb \& Refregier (1997; LR effect) 
point out a systematic effect
due to gravitational lensing of discrete background radio sources
by the clusters, which causes an over-removal
of unresolved radio background thus an underestimate of $\Delta T/T$
of the true SZ effect of the cluster (i.e., a more negative value 
of $\Delta T/T$ than the true value) and subsequently
an underestimate of $H_0$.
They find an underestimate of  $\Delta T/T$ by approximately $5\%$.
We here include the LR effect.
We combine the light dashed curve in Panel (b) of Figure 1
with the Lb effect (assuming to be $5\%$ underestimate of
$\Delta T/T$) and the resulting
distribution of the apparent $H_0$ is shown 
as the solid curve in Figure 2.
Also shown as long-dashed vertical bar is the apparent Hubble
constant when only LR effect is taken into account.
We see that the lensing-coupled pre-SZ CMB fluctuations
produce a fairly broad distribution (with a FWHM of about $18~$km/s/Mpc).
Even if the distribution of $\Delta T/T$ is symmetric, 
that of $H_0$ is {\it not}.
We find that the average and median
values of $H_0$ for the three cases shown 
in Figure 2 (dashed, dotted and solid curves)
are (65.23, 65.00)km/sec/Mpc,
(65.87, 65.00)km/sec/Mpc and
(59.20, 58.50)km/sec/Mpc,
respectively.

It is probably still too early to 
make solid statistical comparisons 
between this model and observations
due to two obvious reasons.
First, the model is perhaps over-simplified.
Second, the observed sample is still too small
[a current total of nine $H_0$ measurements from SZ effect, see Table 2
of Rephaeli (1995)].
For the convenience of the reader, we list the values
of the nine available $H_0$ values from SZ measurements 
(all in units of km/s/Mpc):
$24\pm 11$,
$32^{+19}_{-15}$, 
$38^{+18}_{-16}$,
$40\pm 9$,
$57^{+61}_{-39}$,
$65\pm 25$,
$74^{+29}_{-24}$,
$76^{+22}_{-19}$,
$82^{+35}_{-22}$.
Nevertheless it is quite encouraging that
the effect considered here
provides a reasonable explanation
for the broadness of the observed distribution of $H_0$.

\section{Conclusions}

We show that the background CMB fluctuations, especially when
they are coupled with the gravitational lensing effect by
clusters of galaxies, have one important effect on the 
determination of the Hubble constant through Sunyaev-Zel'dovich
effect of the clusters
(for the adopted set of
characteristic numbers for the cluster, which seem 
fairly realistic compared to those of real clusters of interest):
a broad distribution of the apparent Hubble constant
is produced with a FWHM about $30\%$ of the apparent mean value.
The combination of this effect with other
systematic effects such as the 
Loeb-Refregier Effect seems to provide a reasonable
explanation for the observationally derived values of the Hubble constant
currently available, if the true value of the Hubble constant
is $\sim 65~$km/s/Mpc.
Thus, it becomes possible that the values of $H_0$ measured
by other techniques which generally give a value around $60-80$km/s/Mpc
[e.g., $73\pm 10~$km/s/Mpc ($1\sigma$) from Freedman, Madore, \& Kennicutt 1997
based on HST observations of Cepheids;
$64\pm 6~$km/s/Mpc ($1\sigma$) 
from Riess, Press, \& Kirshner 1996
based on type Ia supernova multicolor light-curve shapes;
$64\pm 13~$km/s/Mpc ($95\%$ confidence level) from Kundic \etal 1997 
based on gravitational lensing time delay measurements;
$70\pm 5~$km/s/Mpc ($1\sigma$) from Giovanelli 1997 
using I-band Tully-Fisher relation;
$81\pm 6~$km/s/Mpc ($1\sigma$) from Tonry 1997 
using surface brightness fluctuations]
be reconciled with the SZ effect determined values of $H_0$.

It may be possible, at least in principle,
that one can use a large sample of SZ measured Hubble constant
to infer the fluctuations of the CMB at the relevant scales,
when the Hubble constant is independently measured to high
accuracy by methods such as that using detached
eclipsing binaries (Paczynski 1997).

\acknowledgments
%I am grateful to ... for discussions.
I like to thank the referee, Dr. Alain Blanchard,
for a number of insightful comments.
The work is supported in part by grants NAG5-2759 and ASC93-18185.

\newpage

\figcaption[Figure 1]{
Panel (a) shows the rms fluctuation of the CMB map 
on a top-hat circle of radius $60^"$
before the gravitational lensing effect 
by the clusters is taken into account (solid histogram).
The light dashed curve is the Gaussian fit with a variance 
of $6.8\times 10^{-6}$ and zero mean.
Panel (b) shows the rms fluctuation of the CMB map 
on a top-hat circle of radius $60^"$
after the gravitational lensing effect 
by the clusters is taken into account (solid histogram).
The light dashed 
curve in the same panel is the Gaussian fit with the same variance 
($1.3\times 10^{-5}$).
\label{fig1}}

\figcaption[Figure 2]{
shows the distribution of $H_0$.
Note that the true Hubble constant is assumed to be $65$km/s/Mpc, as shown
as a vertical solid line at the bottom x-axis.
The dashed and dotted curves indicate
the distributions of the apparent Hubble constant
due to CMB fluctuations {\it without and with}
gravitational lensing effect, respectively.
The solid curve combines the CMB fluctuations and gravitational lensing
effect with the Loeb-Refregier Effect (assuming to be $5\%$ underestimate of
$\Delta T/T$ due to Loeb-Refregier Effect).
Also shown as long-dashed vertical bar is the apparent Hubble
constant when only Loeb-Refregier Effect is taken into account.
\label{fig2}}

\clearpage
\begin{deluxetable}{cccccccc} %{l,r}
\tablewidth{0pt}
\tablenum{1}
\tablecolumns{3}
\tablecaption{CMB fluctuations on arcminute scales} %\label{tab1}}
\tablehead{
\colhead{FWHM} &
\colhead{$(\Delta T/T)_{obs}$ ($10^{-5}$)} &
\colhead{$(\Delta T/T)_{sim}$ ($10^{-5}$)}} 

\startdata
$6^"$  & $7.4\pm 8.1$ & $6.9$ \nl  
$10^"$ & $5.0\pm 5.1$ & $5.1$ \nl 
$18^"$ & $3.3\pm 3.1$ & $3.2$ \nl 
$30^"$ & $2.6\pm 1.9$ & $2.1$ \nl 
$60^"$ & $1.2\pm 1.4$ & $1.2$ \nl
$80^"$ & $1.5\pm 1.5$ & $0.98$ \nl
\enddata
%\tablecomment{TEXT}
%\tablerefs{TEXT}
\end{deluxetable}

\clearpage
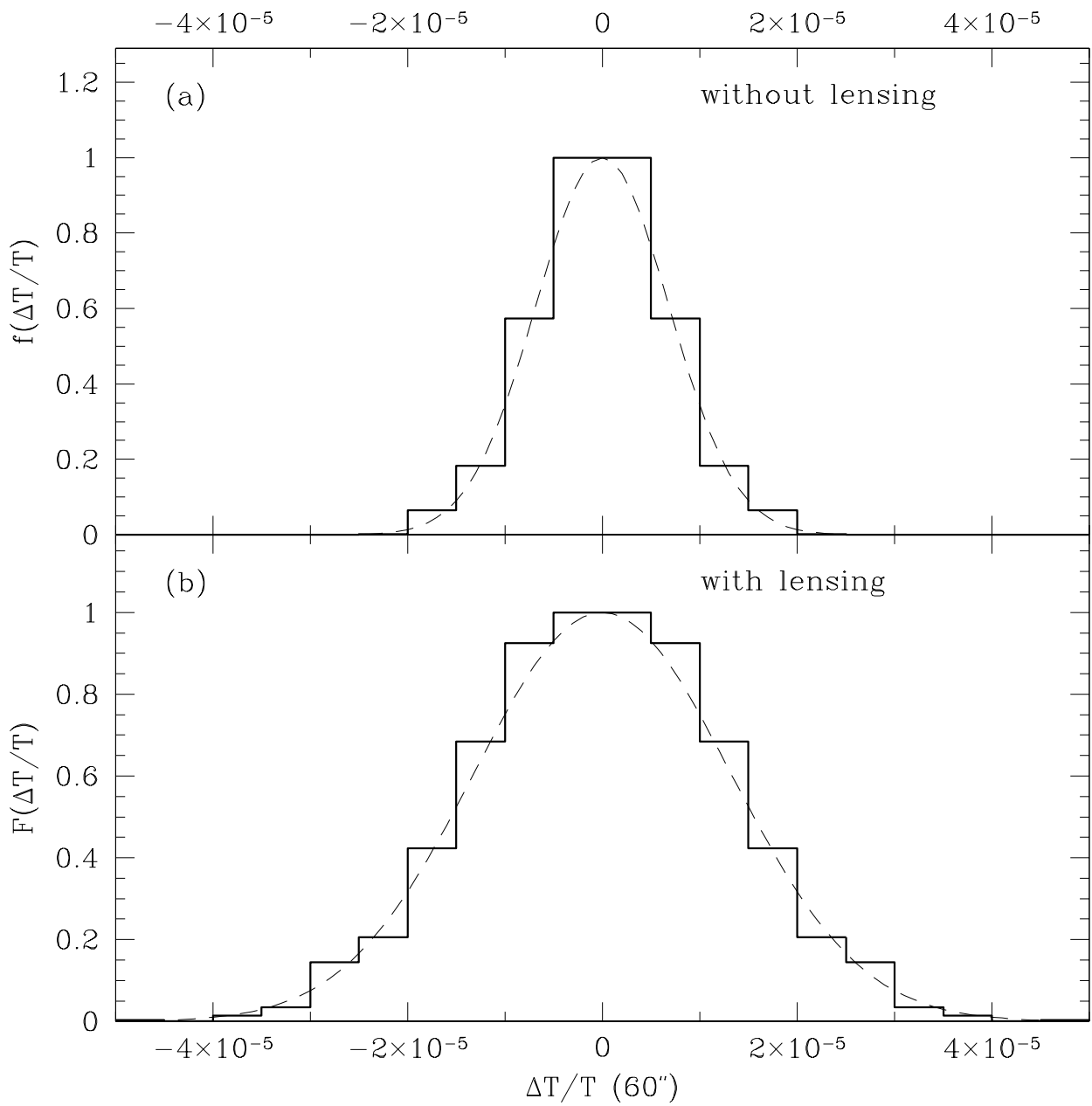
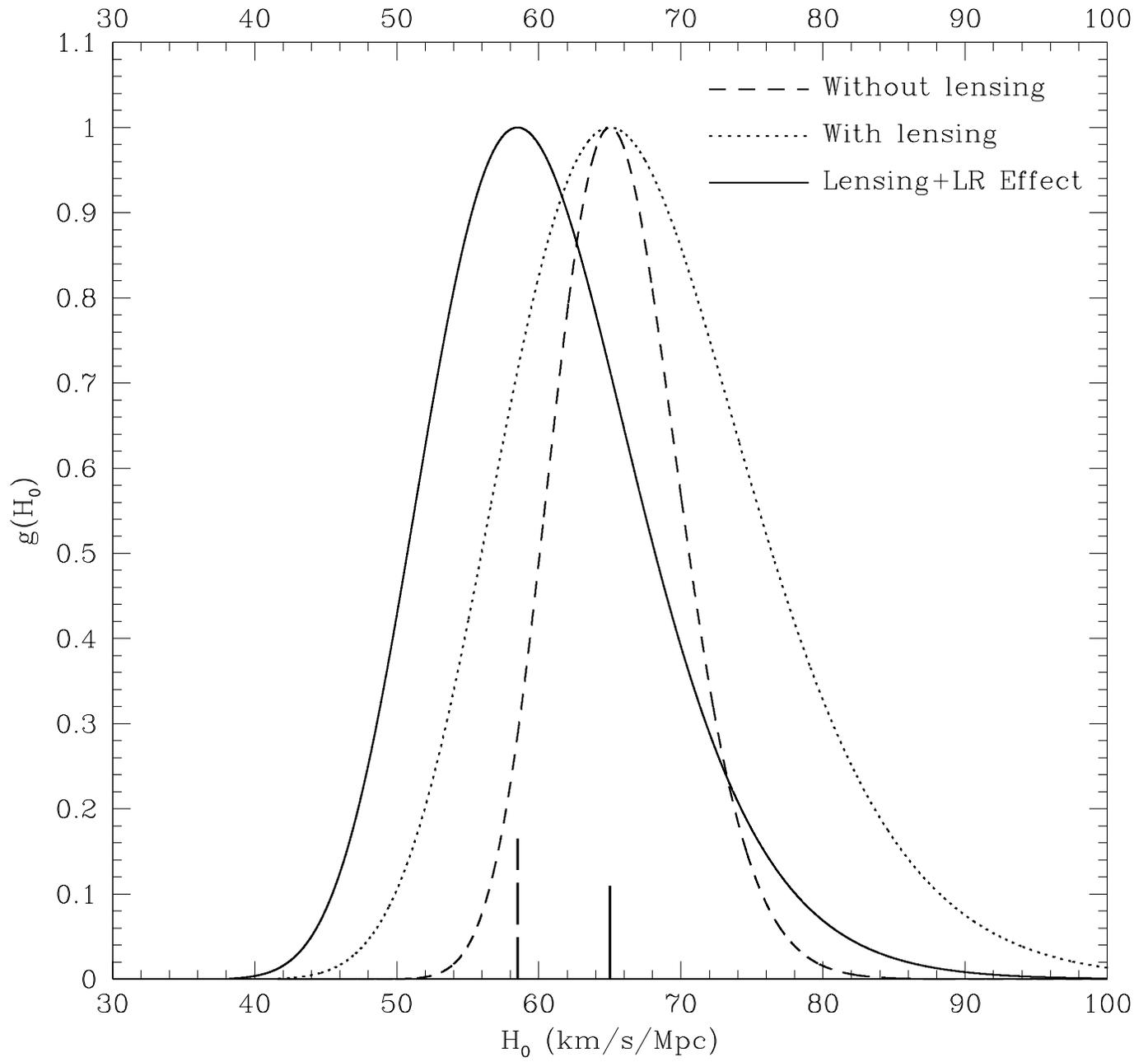
\end{document}